\documentclass[pre,preprintnumbers,amsmath,amssymb,preprint]{revtex4}

\usepackage{graphicx}
\usepackage{bm}

\usepackage{amssymb, dsfont}

\newcommand{\R}{{\mathbf r}} \newcommand{\rr}{{\mathbf r}}
\newcommand{\f}{{\mathbf f}} \newcommand{\F}{{\mathbf F}}
\newcommand{\T}{{\mathbf T}} 
\newcommand{\M}{{\mathbf M}} \newcommand{\K}{{\mathbf K}}
\newcommand{\I}{{\mathds 1}} 
\newcommand{\E}{{\hat{\mathbf e}}} \newcommand{\e}{{\hat{\mathbf e}}}
\newcommand{\V}{{\mathbf V}}

\newcommand{\W}{\mbox{\boldmath$\Omega$}}

\newcommand{\ET}{\mbox{\boldmath$\eta$}}

\newcommand{\beq}{\begin{equation}} \newcommand{\eeq}{\end{equation}}
\newcommand{\bea}{\begin{eqnarray}} \newcommand{\eea}{\end{eqnarray}}

\begin{document}

\title {Spontaneous assembly of colloidal vesicles driven by active swimmers}

\author{Luca Angelani}

\affiliation{ISC-CNR, Institute for Complex Systems, and Dipartimento di Fisica, Universit\`a
Sapienza, Piazzale Aldo Moro 2, I-00185 Rome, Italy}


\begin{abstract} 
We explore the self-assembly process of colloidal structures immersed in active baths.
By considering low-valence particles we numerically investigate the irreversible aggregation
dynamics originated by the presence of run-and-tumble swimmers. We observe the formation of
long closed chains -- vesicles -- densely filled by active swimmers. 
On the one hand the active bath drives the self-assembly of closed colloidal structures,
and on the other hand the vesicles formation fosters the self-trapping of swimmers, 
suggesting new ways 
both to build structured nanomaterials and to trap microorganisms.
\end{abstract}
\maketitle


\section{Introduction}

When immersed in a suspension of active swimmers, like {\it E.coli}
bacteria \cite{Berg},  passive particles and shaped objects experience
effective forces of completely different nature with respect to the
thermal ones. While the latter just produce a Brownian erratic motion
of the passive particles, the active bath has the effect, not only to
enhance the particles diffusivity, but also to give rise to new
phenomena, originated from the out-of-equilibrium  nature of the bath
\cite{Bech_2016,Rei_2017}.  
It has been demonstrated that,
when immersed in active solutions, 
colloidal particles
experience effective attraction forces \cite{Ang_2011}, 
a Casimir-like effect  takes place between parallel plates \cite{reic_2014},
opportunely shaped objects can be spontaneously set into the desired motion \cite{Ang_2009,Rdl_2010,Sok_2010,Ang_2010}, 
polymer chains undergo collapse-expansion dynamics \cite{poly1,poly2,poly3},
flexible membranes manifest shape instability \cite{Nik,Jun},
deformable vesicles exhibit shape-shifting and spontaneous migration \cite{Pao_2016,Tia_2016}.
Moreover, confining structures and boundaries strongly influence swimmers distribution and accumulation,
posing interesting questions about the concept of active pressure 
\cite{Yan_2014,Tak_2014,Fily_2014,Sol_2015,Solon_2015,Smal_2015}.
All these findings highlight the active matter capability to interact
in a non-trivial way  with passive objects, producing new and unusual effects.\\
Up to now only simple or just assembled objects in active solutions
have been considered, such as simple spherical particles, shaped rigid
objects, polymeric-like colloidal chains.  We want now to investigate
the influence of the active bath on the irreversible   aggregation
process of interacting  building blocks.  
The assembly of colloidal particles has been recently
investigated considering intrinsically self-propelled blocks,
such as Janus active Brownian particles or shaped active particles 
\cite{MaCa2016,Mal2018}.
Here, instead, we want to analyze the effect of an
active solution on the aggregation process of simple passive objects.  
In other words we want to answer the question: Is an active
bath able to shape matter during its formation and drive the
assembling of simple passive building blocks into new complex structures?
The recent advances in colloidal manipulation have allowed to synthesize
a huge variety of interacting colloidal objects. 
For example, coating the surface with DNA molecules, one can obtain colloids which exhibit 
reversible and controllable attractions and finite valence \cite{Byr2014,Wang2012}.
We will consider here a system composed
by attractive spherical particles immersed in a bath of run-and-tumble swimmers. 
By considering  particles that can form just
a limited number $k$ of bonds we will study the aggregation process
starting by a uniform distribution of particles in the active
solution.  
We observe the self-assembly of
closed colloidal chains of various lengths   densely filled by active
swimmers. The underlying mechanism at the basis of the observed
vesicles formation is the  persistent character of the swimmers motion
which produces a positive feedback coupling between membrane curvature
and swimmers density.
When small open colloidal chains start to form they are bent by pushing
bacteria until the random attachment of further particles produces a 
long enough chain which eventually folds, forming an {\it active cell},
i.e. a closed colloidal vesicle densely filled  by  swimmers.
On the contrary, the corresponding assembling process in a thermal bath
just produces an ensemble of small clusters composed by few colloidal
particles.  We then claim that active matter is able to drive the
spontaneous formation of complex colloidal structures.

\begin{figure*}[t!]
\includegraphics[width=1\linewidth] {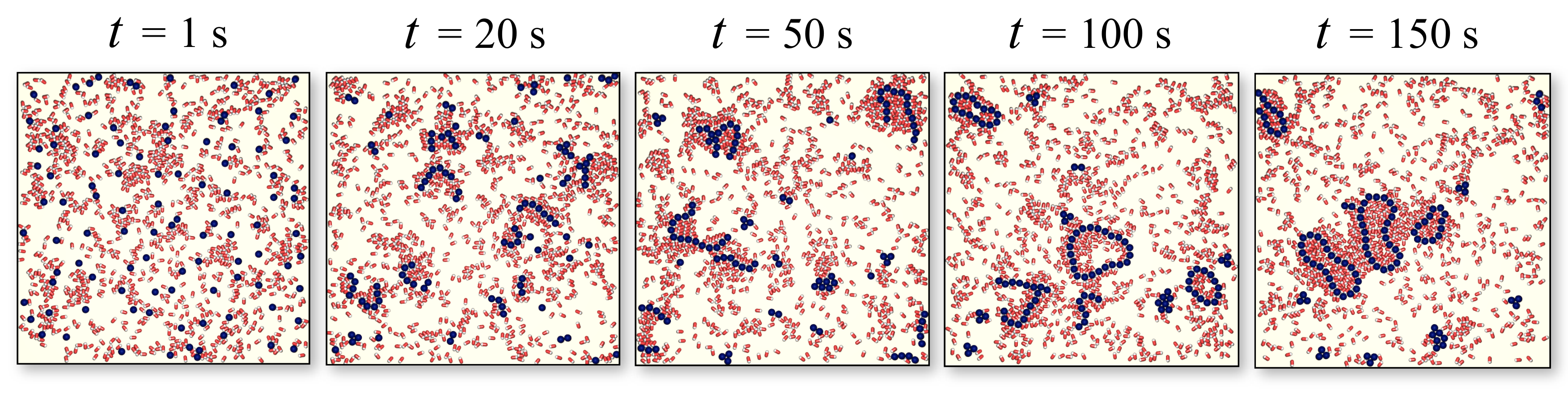}
\caption{\label{fig1}
Colloidal vesicles formation in an active bath.
Simulation snapshots of $N_p=100$ colloidal particles with valence $k=2$ (blue spheres) and 
$N_a=1000$ active swimmers (red-white spherocylinders) at different times,
$t=1, 20, 50, 100, 150$ s. The initial configuration at $t=0$ corresponds 
to a uniform distribution of colloidal particles in the sample.
}
\end{figure*}

\section{ The model}
  
We simulate 
assembling passive particles
immersed in an active bath as follow.
We consider $N_p$ spherical particles  of diameter $d$ (passive system)
in a bath of $N_a$ swimming bacteria of length $l$ and thickness $a$ (active bath).
Each bacterium is represented by a unit vector $\E$, denoting the swimming
direction, and two force-centers along it located at
${\bf r} \pm \e \ l/4$ (we consider swimmer aspect ratio $a/l =1/2$).
Run-and-tumble dynamics are considered for the swimmers: 
a self-propelling force acts along the direction $\E$ and
a random reorientation of the swimming direction
takes place with a probability per unit time $1/\tau_{run}$.
The tumble event lasts $\tau_{tumble}$.
Excluded volume interactions among swimmers and particles are
described by pair-repulsive forces acting between force-centers of
different organisms.  Swimmer-particle and particle-particle
interactions are  designed to have the same stiffness of
swimmer-swimmer interaction.  This is achieved by considering
particles force-centers located  $a/2$ behind the particle surface. 
The force on $i$-th micro-organism due to its
self-propulsion and the mechanical interactions  is
\begin{equation}
\F_i = f_0 \E (1-\theta_i) + \sum_{j\neq i} \sum_{\beta,\gamma}
\ \f(\R_i^\beta-\R_j^\gamma) + \sum_{p,\beta} \ \f(\R_{p,i}^\beta)
\end{equation}
where $f_0$ is the self-propulsion force, $\theta_i = 0,1$ respectively in the
run and tumble state, index $j$ runs over swimmers, 
indexes  $\beta, \gamma =1,2$ run over the swimmers force-centers, 
index $p$ runs over particles,
$\R_{p,i}^\beta=\R_i^\beta-\R_p - [(d-a)/2] (\R_i^\beta-\R_p)/|\R_i^\beta-\R_p|$ with $\R_p$ the particle
position vector 
and $\f$ is a
repulsive force (truncated Lennard-Jones),
${\bf f}(\rr)={\bf {\hat r}} f(r)$
with $f(r)=f_{rep} [(a/r)^{13} - (a/r)^{7}]$ 
for $r<a$ and 0 otherwise.  
The corresponding torque on $i$-th swimmer is given by the
expression:
\begin{eqnarray}
\T_i &=&  \T_r \theta_i + \sum_{j\neq i} \sum_{\beta,\gamma}
\ (-1)^{\beta}\frac{l}{4}  \f(\R_i^\beta-\R_j^\gamma) + \nonumber
\\ &+& \sum_{p,\beta}  \ (-1)^{\beta}\frac{l}{4}  \f(\R_{p,i}^\beta)
\end{eqnarray}
where $\T_r$ is a random torque acting during the tumbling event.  

As passive particles we consider short-range attractive colloids.
Each particle is considered to have a finite valence $k$, i.e. a maximum number of bonds allowed.
Real attractive colloids, such as DNA-coated particles, 
have localized sticky points on their surface, resulting in anisotropic interactions between them. 
However, it has been shown that the bond flexibility can be controlled, 
for example by choosing different DNA linker lengths in the case of coated colloids \cite{Byr2014}.
In the present numerical study we assume maximum bond flexibility 
modeled by isotropic interactions between particles \cite{Mark2012}.
The total force on $p$-th passive particle is given by
\begin{equation}
\F_p =  \sum_{q\neq p} \ [\f(\R_{p,q}) + c_{p,q} \f_{bond}(\R_{p,q}) ]
- \sum_{i,\beta} \ \f(\R_{p,i}^\beta)
\label{Fp}
\end{equation}
where the first sum is over particles and the second one over swimmers,
$\R_{p,q}=\R_q-\R_p - (d-a) (\R_q-\R_p)/|\R_q-\R_p|$, ${\bf f}$ is the repulsive force 
and ${\bf f}_{bond}$ the attractive one.
In order to simulate the assembling process of the building blocks 
we consider that each passive particle can form at most $k$ bonds with other particles.
Initially particles are not-bonded, i.e. $c_{p,q}=0$.
As soon as two particles become into contact (their distance is less than $a$)
and none of the two particles has reached the maximum allowed number $k$ of bonds, 
a bonding force sets in for the rest of the simulation, $c_{p,q}=1$.
In such a way we simulate the irreversible aggregation process of colloidal particles.
The form of the bonding force is chosen to be radial and similar to the steric repulsive force,
specularly reflected with respect to $d$, i.e. 
$\f_{bond}(\rr) = - {\bf{\hat r}} f_{bond}(r)$ with
$f_{bond}(r) = f(2a-r)$ for $r>a$ and 0 otherwise.
In the present study we neglect hydrodynamic interactions,
following previous similar investigations where their contribution has been demonstrated
to have negligible or little  effects on the obtained results \cite{Ang_2011,Ang_2009,Rdl_2010,Ang_CPC}.

\begin{figure*}[t!]
\includegraphics[width=1\linewidth] {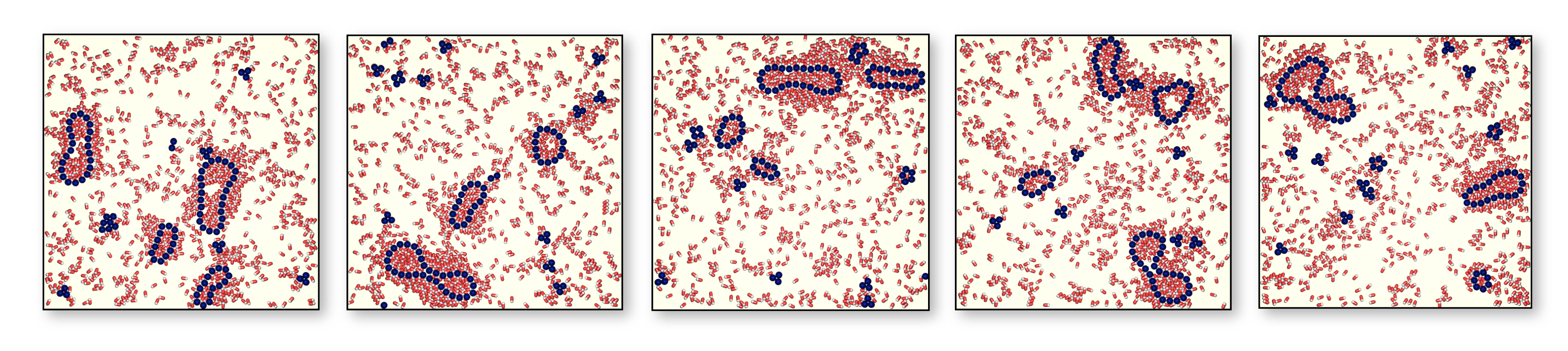}
\caption{\label{fig2}
Examples of final vesicles conformation in independent samples.
Simulation snapshots of $N_p=100$ colloidal particles with valence $k=2$ and 
$N_a=1000$ swimmers from five independent simulation runs at long times.
}
\end{figure*}

The equations of motion for the swimmers in the low Reynold numbers regime  are
$\V_i = \M_i \cdot \F_i$ and $\W_i = \K_i \cdot \T_i$,
where $\V$ and $\W$ are translational and angular swimmers velocities,
$\M=m_{||}\e \e+m_{\perp}\left(\I-\e \e \right)$ and $\K=k_{||}\e \e
+k_{\perp}\left(\I-\e \e \right)$ their translational and rotational
mobility matrices, $\F$ and $\T$ the force and torque \cite{kim}.
The equations of motion for the passive spherical particles are  
$\V_p = M_p \ \F_p$,
where $M_p$ is the particles translational mobility.
The equations of motion are numerically integrated  for 
$2 \times 10^7$
time steps by using the Runge-Kutta method \cite{numrec}, with time step
$5 \times 10^{-4}$ 
(unless differently specified quantities are expressed in
internal units:  $l$ for length, $m_{||}$ for mobility, $f_0$ for
force; physical units are obtained setting 
$l=3\ \mu$m, $m_{||}=59\ \mu$m pN$^{-1}$ s$^{-1}$, $f_0=0.51$ pN, appropriate 
for {\it E.coli} bacteria).
We set mobility values for swimmers  $m_{\perp}=0.87$
and $k_{\perp}=4.8$,  obtained assuming a prolate spheroids shape of
bacteria \cite{kim}, and $\tau_{run}=10$, $\tau_{tumble}=1$.
We consider spherical particles of diameter $d=1$ 
and mobility value $M_p=0.6$, appropriate for spherical objects.
We will also investigate the case of bigger particles, with diameter $d=2$.
The strength of the repulsive force is chosen $f_{rep}=1$.
Simulations of $N_a=1000$ and $N_a=600$ swimmers and $N_p=100$ particles are performed
in a 2D box of length $L=40$ ($L=120\ \mu$m in physical unit)
with periodic boundary conditions.
The bacteria and particles densities are respectively 
$\rho_b = 6.25 \times 10^{-1}$ ($\rho_b = 6.94 \times 10^{-2} \mu$m$^{-2}$ in physical unit) 
and $\rho_p=6.25 \times 10^{-2}$ ($\rho_p = 6.94 \times 10^{-3} \mu$m$^{-2}$ in physical unit).

\noindent 
For comparison we also simulate passive colloidal particles in a thermal bath, 
considering the particles equations of motion 
$\V_p = M_p \ \F_p + \ET$, where $\F_p$ is given by Eq. (\ref{Fp})
without the last term (related to the active bath)
and $\ET$ is a Gaussian white noise which
satisfies $\langle \eta_\alpha(t) \rangle = 0$ and
$\langle \eta_\alpha(t)  \eta_\beta(t')  \rangle = \delta_{\alpha \beta}
2 D \delta(t-t')$ with $\alpha,\beta = x,y$.
The equations of motion are numerically integrated for 
$2 \times 10^8$ time steps.
The diffusion coefficient  $D=M_p k_B T$ is set to the value
$1.6 \times 10^{-3}$ (in reduced units), which in physical units 
corresponds to  $0.14\ \mu$m$^2 /$s, appropriate
for a colloidal particle of radius $1.5 \ \mu$m in water at room temperature $T=293.15$ K.
To elucidate the role of steric repulsion between colloids and swimmers in the ring formation process
we also consider the case of dead bacteria, simulating passive colloids immersed in a bath 
of dead elongated cells, both agitated by thermal fluctuations.

\begin{figure*}[!]
\includegraphics[width=1\linewidth] {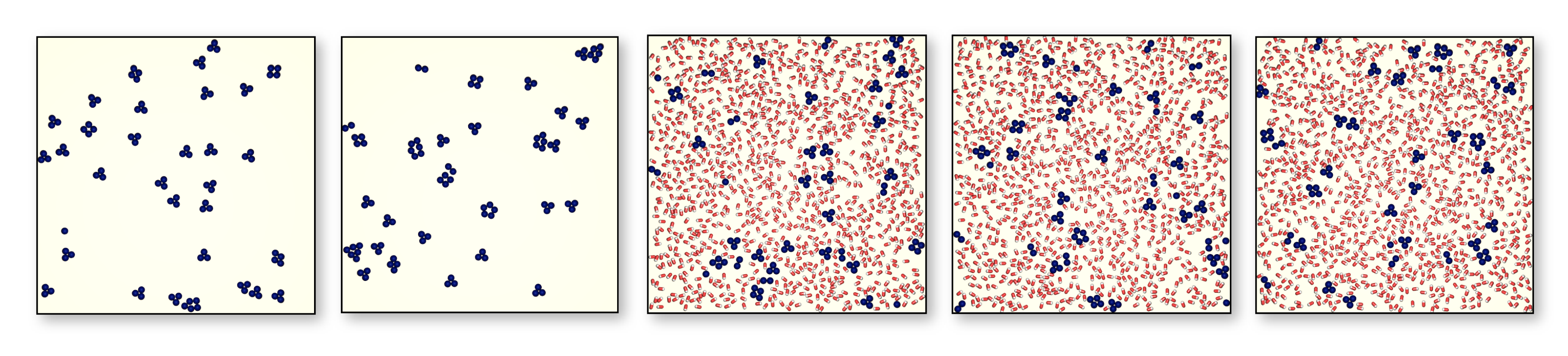}
\caption{\label{fig3}
Simulation snapshots of $N_p=100$ colloidal particles 
with valence $k=2$ immersed in a thermal bath.
The two panels on the left refer to the case of thermal colloidal particles without bacteria.
The three panels on the right are obtained considering colloids immersed in a bath of dead bacteria.
}
\end{figure*}

\section{Results}
  
We consider $N_p$ sticky colloidal particles of low-valence $k$ which are initially uniformly distributed 
in a square area of length $L$ filled by an active bath of $N_a$ swimmers. 
We analyze the irreversible aggregation dynamics of the particles over time looking at the spatial
configurations and structures of colloids and swimmers.\\
We first consider the case of particles valence $k=2$,
i.e. each colloidal particle can form at most two bonds with other particles.
In Fig.\ref{fig1} snapshots from a simulation run are shown
at different times. We observe a clear tendency of particles to form 
polymer-like bended structures which eventually form
closed chains (vesicles) densely filled by active swimmers. 
At long time,
when all the particles have saturated all their allowed bonds ($k=2$ in this case),
the emerging structures consist of few large membranes with active matter inside 
plus isolated small clusters of few particles. Once assembled, such active cells,
formed by a colloidal membrane plus active swimmers inside,
perform shape-shifting dynamics and spontaneous migration along polarization directions,
as already observed in a recent work where  just assembled vesicles
filled up by construction with motile bacteria have been analyzed \cite{Pao_2016}. 
The relevant point here is that such complex structures emerge spontaneously from
the simulation and there is not a priori external constraints imposing the system
to form closed rings surrounding active swimmers. The formation of these 
assembled structures is driven by the swimmers itself, which
cooperate to induce a self-caging effect. This process is a direct consequence
of the active matter property to accumulate at convex boundaries,
where swimmers align their self-propelled direction with the boundary normal
and exert a pressure proportional to the local curvature. This pressure,
in the case of flexible boundaries, has the effect to increase the 
boundary curvature and then to further enhance the swimmers accumulation \cite{Pao_2016}.
The curvature-density feedback mechanism produces high-curvature folded chains
pushed by groups of swimmers (see Fig.\ref{fig1}).
Such curved open chains eventually end up in closed rings of different lengths.
In Fig.\ref{fig2} examples of five different configurations obtained from independent simulation runs
at long times are shown. It is evident the clear tendency to form colloidal {\it cells}
filled by swimmers.
The typical length of the longer vesicles is about $20-30$ particles. The average clusters length 
at long time, considering all the clusters and  averaging over $20$ independent samples, 
is $l_{cl}\simeq 8$.\\
To highlight the relevant role of active swimmers in the vesicles formation we have also
considered the aggregation process of attractive colloidal particles in 
a thermal bath. 
Firstly we have considered the case of purely thermal colloidal particles, in the absence of 
swimmers.
In Fig.\ref{fig3} (the two panels on the left) we report snapshots of particles configurations 
obtained in independent simulation runs at long times.
One observes, in this case, the formation of small clusters made of few particles
and no long closed chains emerge during the simulation.
Occasionally one observes closed chains made of up $10$ particles, but the majority 
of the clusters have length less than $5$. The average clusters length 
at long time is in this case $l_{cl}\simeq 3$.
To elucidate the role of excluding volume effects of the swimmers in the vesicles formation 
we have also
simulated colloidal particles immersed in a bath of dead bacteria.
Three typical configurations obtained in the simulations are reported in Fig.\ref{fig3} (right panels).
No significant differences are observed in the small clusters formation made of few particles 
with respect to the previous case of absence of bacteria, thus indicating that the relevant 
ingredient for the large clusters formation is the active nature of the bath.
It is worth noting that, by construction, the studied model involves forming closed structures.
However, these turn out to be small clusters of few particles in the thermal case,
while large vesicles appear in the active solution.
We can then claim that the activity of the bath has the effect of inhibiting 
the formation of many small colloidal clusters
and fosters the elongation of the chains until they coalesce into closed rings.
\\
\begin{figure}[t!]
\includegraphics[width=0.8\linewidth] {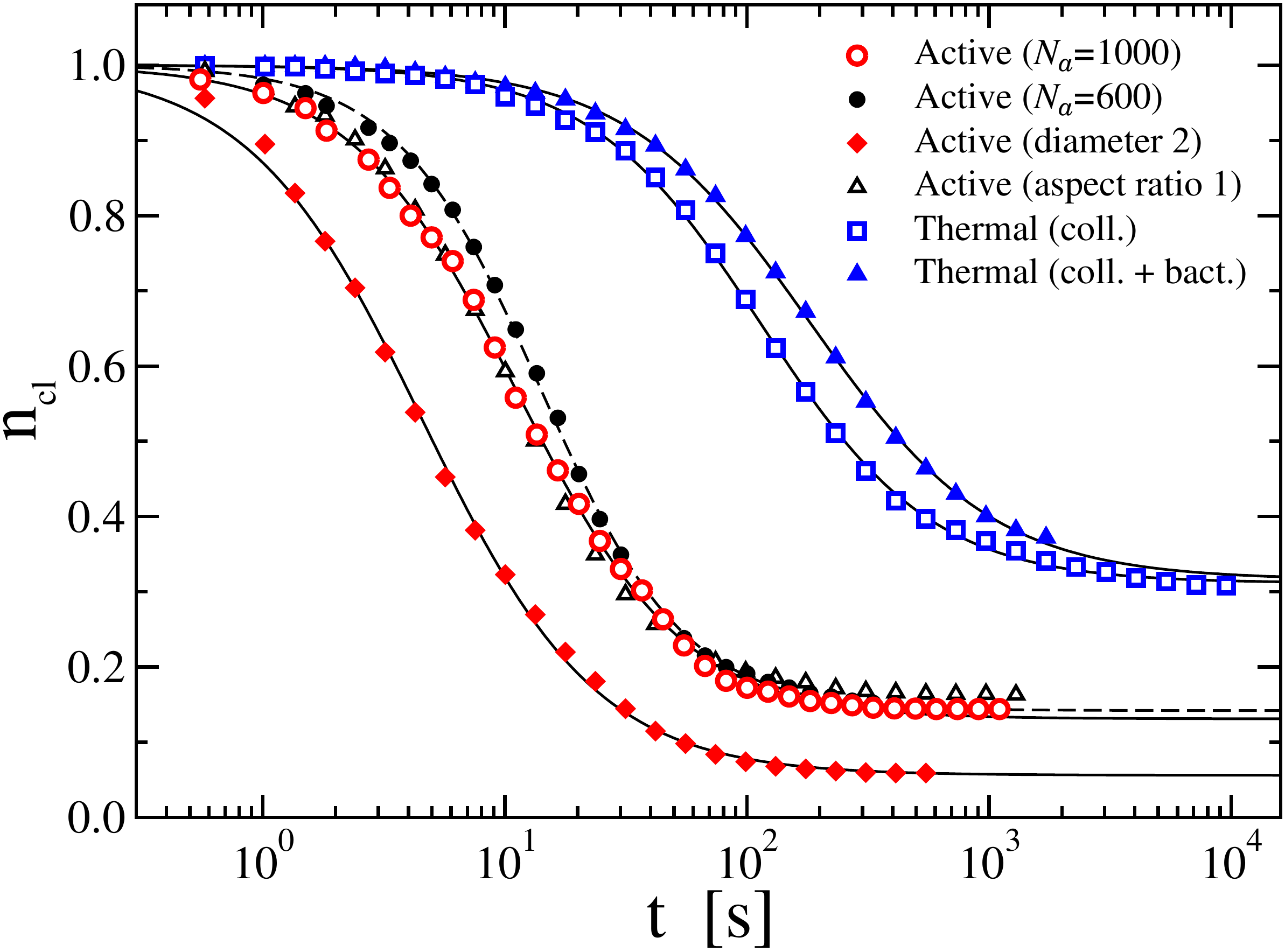}  
\caption{\label{fig4}
Fractional number of colloidal clusters $n_{cl}=N_{cl}(t)/N_{cl}(0)$
as a function of time in the case of active  and thermal bath.
$N_{cl}(t)$ is the number of clusters at time $t$ and $N_{cl}(0)=N_p$
is the initial number of clusters, corresponding to uniformly distributed disconnected particles.
Data are referred to the case of $N_p=100$ colloidal particles with valence $k=2$.
In the case of active baths we report data for the main system (swimmers number $N_a=1000$, 
swimmers aspect ratio $1/2$, colloids diameter $d=1$) and also for lower swimmers density 
($N_a=600$), bigger particles size ($d=2$) and swimmers aspect ratio $1$.
Two different thermal baths are considered: the case of thermal colloids without bacteria
and the case of both thermal colloids and dead bacteria.
Lines are fits with a log-logistic like function 
$n_\infty + (1-n_\infty) / [1+(t/t_0)^\alpha]$:
$n_\infty \simeq 0.134$, $\alpha \simeq 1.27$, $t_0 \simeq 11.2$ s
(active bath, $N_a=1000$); 
$n_\infty \simeq 0.142$, $\alpha \simeq 1.46$, $t_0 \simeq 14.0$ s
(active bath, $N_a=600$); 
$n_\infty \simeq 0.056$, $\alpha \simeq 1.21$, $t_0 \simeq 4.57$ s
(active bath, $N_a=1000$, $d=2$); 
$n_\infty \simeq 0.312$, $\alpha \simeq 1.23$, $t_0 \simeq 114$ s
(thermal bath);
$n_\infty \simeq 0.316$, $\alpha \simeq 1.15$, $t_0 \simeq 182$ s
(thermal bath and dead bacteria).
}
\end{figure}
To quantify the cluster formation process we calculate the time-dependence of the 
fractional number of clusters $n_{cl}(t)=N_{cl}(t)/N_{cl}(0)$, where $N_{cl}(t)$ is the number of clusters
at time $t$. A cluster is defined as a structure composed by connected particles.
At initial time $t=0$ the  particles are uniformly distributed in the box area and each
cluster is composed by a single particle, then $N_{cl}(0)=N_p$.
We have $n_{cl}=1$ at initial time when all the particles are not-bonded, 
and $n_{cl}=1/N_p$ when there is a single global structure made of $N_p$ particles.
In Fig.\ref{fig4} the quantity $n_{cl}$, averaged over $20$ independent runs, is reported as a
function of time for the cases of active 
and thermal baths.
For active baths we have also considered the cases of lower swimmers density ($N_a=600$) 
and bigger size of colloids ($d=2$) with respect to the investigated values 
$N_a=1000$ and $d=1$.
The asymptotic fractional number of clusters is about 0.14, 
corresponding to an average number of clusters $N_{cl}=14$,
in the case of active baths ($N_a=1000$),
and 0.31 ($N_{cl}=31$) 
in the case of the two different thermal baths, i.e. with and without dead bacteria.
The lower number of clusters in the active case corresponds to the presence
of longer closed chains.
No significant differences are observed by varying the swimmers density,
but a slightly longer relaxation time due to the lower number of pushing bacteria.
Considering a bigger size of the particles one observes that the average number of cluster decreases, 
indicating the tendency to form larger clusters, maybe due to the 
higher colloidal area packing fraction enhancing collision probability.
It is worth noting that the shape of the active particles has a little effect 
on the vesicles formation. Indeed, by simulating spherical swimmers (aspect ratio 1)
the resulted curve of fractional clusters number as a function of time superimposes to the one obtained
with elongated bacteria (see Fig.\ref{fig4}).

\begin{figure*}[h]
\includegraphics[width=1\linewidth] {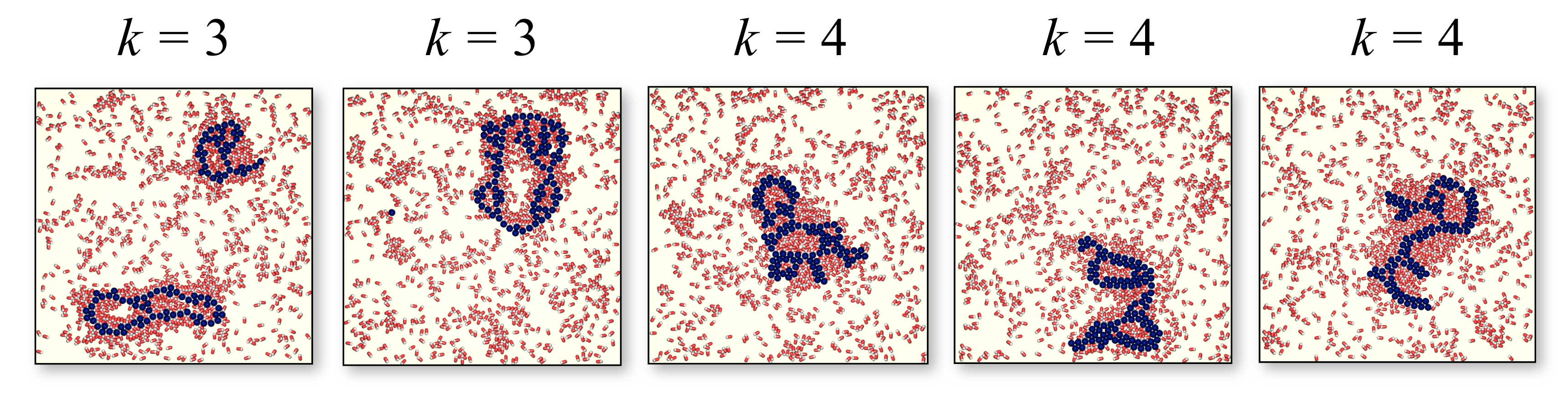}
\caption{\label{fig5}
Examples of colloidal structures with high valence colloids.
Simulation snapshots of $N_p=100$ colloidal particles with valence number $k=3$
and $k=4$  and 
$N_a=1000$ swimmers from independent simulation runs at long times.
}
\end{figure*}

We now consider the cases of interacting colloidal particles with 
higher valence number, $k>2$.
We have simulated systems with particle valence $k=3$ and $k=4$.
Example of final configurations obtained at long time are reported in Fig.\ref{fig5}.
As in the previous case we observe closed colloidal structures
filled by active swimmers. However, in this case the colloidal aggregates are
more complex and characterized by protrusions, intersections, multiple connected vesicles,
and, moreover, all the particles coalesce into just one (or few) final global structure.

The reported numerical results lead the way to new experimental investigations.
One can conceive experiments in which low valence particles, such as 
patchy colloids, i.e.  colloidal particles with a finite number of attractive sites in 
their surface \cite{bia1,Jia,bia2,ruz},  are immersed in active suspensions of {\it E.coli} bacteria or artificial
self-propelled particles. However, in the case of patchy particles, at variance with the present study
in which spherical interactions have been considered, 
anisotropic interactions could lead to more rigid structures and there could be some
differences in the final colloidal aggregate structures. 
It remains an open interesting question to elucidate
the relationship between the degree of anisotropy of the colloidal interactions and 
the property of the emerging structures.
Other interesting open questions regard the role of hydrodynamics in 
the vesicles formation and how the vesicles size is related to the bath activity,
or, in other wards, how the vesicles formation process varies with the P\'eclet number, i.e. the ratio between 
active and thermal forces.

\section{Conclusions}
 
Active baths of self-propelled swimmers are able to forge passive matter
and to drive the formation of complex structures. 
We have demonstrated, through numerical simulations, that low-valence colloidal
particles spontaneously form long ring-like structures when immersed in a bath 
of run-and-tumble swimmers. 
The same passive colloidal system in a thermal bath 
gives rise to many small clusters composed by few particles.
The active bath is then responsible for the formation of such colloidal
vesicles, originated by the persistent character of the simmers motion which 
facilitate the formation of long colloidal chains and foster their folding
in closed membranes.
Moreover, such colloidal vesicles turn out to be densely filled by
active swimmers, indicating that the self-assembly of colloids
also induces the self-trapping of swimmers. 
Combining passive matter, such as colloidal low-valence particles,
with active systems, e.g. {\it E.coli} bacteria or self-propelled Janus particles,
promotes the emergence of complex structures, suggesting new ways to control
nanomaterials formation and microorganisms confinement.





\end{document}